\documentclass[aps, prl, amsfonts, groupedaddress,twocolumn]{revtex4-2}
\usepackage{latexsym}
\usepackage{bm}
\usepackage{graphicx}        
\usepackage{amssymb}
\usepackage{amsmath}
\usepackage{units}

\usepackage[usenames,dvipsnames]{xcolor}
\usepackage[colorlinks=true, urlcolor=urlcolor, linkcolor=linkcolor, bookmarks=true, citecolor=blue, breaklinks=true]{hyperref}

\colorlet{notecolor2}{olive}
\definecolor{notecolor}{RGB}{150,20,40}
\definecolor{linkcolor}{RGB}{20,90,15}
\definecolor{urlcolor}{RGB}{150,20,40}

\newcommand{\Up}{U_\mathrm{p}}

\newcommand{\Efield}{\mathcal{E}}

\begin{document}

\title{High-order phase-dependent asymmetry in the above-threshold ionization plateau}

\author{M. K\"ubel}\email{matthias.kuebel@uni-jena.de}
\author{P. Wustelt}
\author{Y. Zhang}
\author{S. Skruszewicz}
\author{D. Hoff}
\author{D. W\"urzler}
\author{H. Kang}
\author{D. Zille}
\author{D. Adolph}
\author{A. M. Sayler}\altaffiliation{Present address: Benedictine College, Atchison, KS, USA}
\author{G. G. Paulus}
\affiliation{Institute of Optics and Quantum Electronics, Max-Wien-Platz 1, D-07743 Jena, Germany}
\affiliation{Helmholtz Institute Jena, Fröbelstieg 3, D-07743 Jena, Germany}

\author{M. Dumergue}
\author{A. Nayak}
\author{R. Flender}
\author{L. Haizer}
\author{M. Kurucz}
\author{B. Kiss}
\author{S. K\"uhn}
\affiliation{ELI-ALPS, ELI-HU Non-Profit Ltd., Wolfgang Sandner utca 3., Szeged, H-6728, Hungary}

\author{B. Feti\'{c}}
\author{D. B. Milo\v{s}evi\'{c}}

\affiliation{Faculty of Science, University of Sarajevo, Zmaja od Bosne 35, 71000 Sarajevo, Bosnia and Herzegovina}
\affiliation{Academy of Sciences and Arts of Bosnia and Herzegovina,
Bistrik 7, 71000 Sarajevo, Bosnia and Herzegovina}


%
\date{\today}

\begin{abstract}

Above-threshold ionization spectra from cesium are measured as a function of the carrier-envelope phase (CEP) using laser pulses centered at 3.1 $\mu$m wavelength. The directional asymmetry in the energy spectra of backscattered electrons oscillates three times, rather than once, as the CEP is changed from $0$ to $2\pi$. Using the improved strong-field approximation, we show that the unusual behavior arises from the interference of few quantum orbits. We discuss the conditions for observing the high-order CEP dependence, and draw an analogy with time-domain holography with electron wave packets.

\end{abstract}

\maketitle
Controlling electron motion using tailored laser fields is a central goal of strong-field and attosecond physics. This includes the motion of quasi-free electrons, underlying attosecond pulse generation \cite{Baltuska2003} and encoding ultrafast temporal information \cite{Kienberger2004}; valence electrons in molecules \cite{Kling2006}, determining the outcome of chemical reactions; or ultrafast currents in solids \cite{Schiffrin2013}.
Various control techniques have been proposed and implemented, including attosecond pulses \cite{Goulielmakis2004}, multi-color pump-probe schemes \cite{Kuebel2017, Kuebel2020}, and polarization-shaped laser pulses \cite{Mashiko2008,Kfir2015}.

One of the most fundamental approaches, however, relies on the use of a few-cycle laser pulse, $\Efield (t;\phi) = \Efield_0(t) \cos{\left(\omega t + \phi\right)}$, with a controlled or known carrier-envelope  phase (CEP) $\phi$, and an envelope function whose duration is comparable to an optical cycle $T=2\pi / \omega$. This approach has been used to control various processes in atoms, molecules, and solids. The prototypical CEP effect consists in an asymmetry in the above-threshold ionization (ATI) spectra of photoelectrons emitted into opposite directions along the laser polarization \cite{Paulus2001_Nature}. The sinusoidal oscillations of the CEP-dependent asymmetry, $A(\phi) \propto \sin(\phi+\phi_0)$, represents the basis for measuring the CEP using the stereo-ATI technique \cite{Paulus2003,Rathje2012}, or other approaches \cite{Paasch-Colberg2014, Kubullek2020}. Here, $\phi_0$ is a phase offset, known as the phase-of-the-phase \cite{Skruszewicz2015}, that generally depends on the electron drift momentum. 

The CEP-dependence of ATI can be understood on a qualitative level by simple symmetry considerations: a CEP-stable few-cycle laser pulse has broken inversion symmetry. Since the drift momenta of photoelectrons are determined directly by the electric field evolution, the asymmetry of the pulse is transferred onto the motion of the field-driven electrons. As the field asymmetry is maximized for $\phi=0,~\pi$, and the field is symmetric for $\phi=\pi/2,~3\pi/2$, one expects the aforementioned sinusoidal oscillation for the photoelectron asymmetry, in accordance with experimental results. CEP effects are particularly pronounced for  recollision processes \cite{Corkum1993}, including high-order ATI \cite{Paulus2003}, non-sequential double ionization \cite{Liu2004, Bergues2015}, and high-harmonic generation (HHG) \cite{Baltuska2003}.

CEP effects with periodicity of $\pi$ rather than $2\pi$, i.e., with periodicity parameter $m=2$ instead of $m=1$ are observed for the photon yields from HHG \cite{Baltuska2003, Hollinger2020} and for total yields of double ionization \cite{Johnson2011, Kuebel2012} or fragmentation \cite{Xie2012}. The doubled periodicity is consistent with the above symmetry considerations since total yields are insensitive to the direction of the field but are affected by the modulation of the instantaneous intensity resulting from varying the CEP. 
The general theory of CEP effects \cite{Roudnev2007} predicts oscillations with periodicity parameter $m$ to result from the interference of two pathways that involve the absorption of $n$ and $n+m$ photons, respectively. However, experimental evidence for $m>2$ has been lacking.

Here, we present ATI measurements of Cs using CEP-stable few-cycle laser pulses with a central wavelength of 3100\,nm. At the laser intensity of $4\,\mathrm{TW/cm^2}$, we observe an unusual CEP-dependent asymmetry with $m=3$ in the ATI plateau region around $6\Up$, where $\Up$ is the ponderomotive potential. For lower intensity values, the CEP-dependent asymmetry exhibits the usual behavior with $m=1$. 
The experimental results are interpreted using the improved strong-field approximation (ISFA) and using the saddle-point method \cite{Milosevic2006}.
We show that the observed fast oscillations in the CEP-dependent asymmetry are due to the interference of a few quantum orbits \cite{Milosevic2005} which are modulated by the CEP.  We discuss the conditions under which high-order CEP-dependent asymmetries are observable in ATI. 

The experiments are conducted using the mid-infrared (MIR) laser at the Extreme Light Infrastructure Attosecond Light Pulse Source (ELI-ALPS). It provides intense ultrashort ($\sim 42\,$fs) laser pulses centered at $\lambda = 3100\,$nm at a repetition rate of $100\,$kHz \cite{Thire18}. The laser pulses are post-compressed to a pulse duration (full width at half-maximum of the intensity envelope) of $\tau_p = 31\,$fs (corresponding to three optical cycles) using the method described in Ref.~\cite{Kurucz2020} \footnote{The increased pulse duration with respect to the $\sim 24\,$fs reported in Ref.~\cite{Kurucz2020} is attributed to the presence of high-order dispersion caused by using a $2\,$mm rather than a $1\,$mm thick Si plate in the beam path.}.
The CEP of the laser pulses has an excellent stability with a jitter of only $82\,$mrad root mean square \cite{Kurucz2020}. In our experiments, the CEP is controlled with a pair of BaF$_2$ wedges.  

The linearly polarized laser pulses are focused with a thin CaF$_2$ lens ($f = 200\,$mm) into a vacuum chamber with base pressure of $10^{-7}\,$mbar, where they intersect an atomic beam of Cs, produced by evaporating Cs at $\sim 100^\circ$C. The ionization potential of Cs ($3.9\,$eV) is much lower than that of the background gas (mostly N$_2$, O$_2$, H$_2$O). Hence, contamination of the measured photoelectron signal by ionization of background gas is negligible at the relatively low intensities ($I<10^{13}\,\mathrm{TW/cm^2}$) used in our experiments. The photoelectrons generated in the laser focus are detected using two pairs of multichannel plates detectors of 25\,mm diameter, placed at the ends of two $50\,$cm long drift tubes along the laser polarization. The geometry supports a detection angle of $\sim 3^\circ$, such that essentially only directly forward or directly backward scattered electrons are detected.  Using a pair of detectors on either side of the laser polarization (i.e., the stereo-ATI technique) helps suppressing the effect of laser intensity fluctuations on the recorded CEP-dependent ATI spectra.

\begin{figure}
\includegraphics[width=0.48\textwidth]{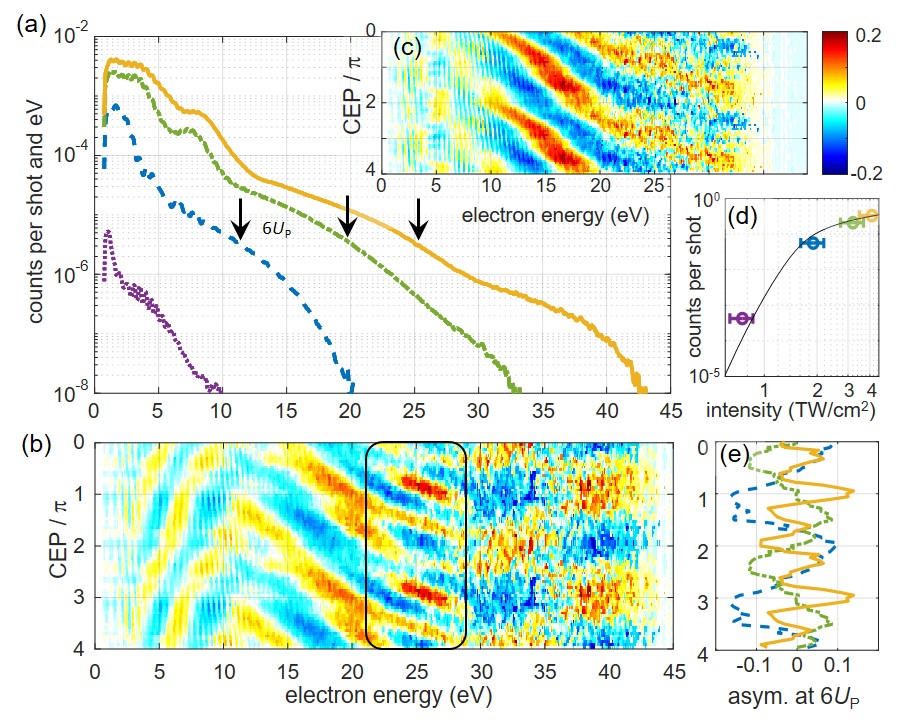}
\caption{Experimental results for ATI of Cs by linearly polarized few-cycle mid-IR laser pulses ($\tau_p = 31\,$fs, $\lambda = 3100\,$nm). (a) CEP-averaged photoelectron spectra measured for intensity values of approximately $0.7\,\mathrm{TW/cm^2}$ (purple dotted line), $1.8\,\mathrm{TW/cm^2}$  (blue dashed line), $3.2\,\mathrm{TW/cm^2}$ (green dashed-dotted line), and $4.1\,\mathrm{TW/cm^2}$ (yellow solid line), respectively. (b,c) Measured CEP-dependent asymmetry parameter for 4.1 and 3.2$\,\mathrm{TW/cm^2}$, respectively. The black box in (b) marks the trifurcation of the asymmetry, which is absent in (c). (d) The average number of electrons detected per laser shot is compared to predictions based on the PPT tunneling rate (solid line). The error bars depict an estimated 15\% uncertainty in the intensity determination. Panel (e) presents the CEP-dependent asymmetries at electron energies of $6\Up$, as marked by the black arrows in (a). The high-order, $m=3$, asymmetry oscillation in the case of the highest intensity is clearly visible and distinct from the usual $m=1$ oscillations observed for the lower intensity values.} 
 \label{fig1}
\end{figure}

Photoelectron spectra for ATI of Cs ionized by few-cycle pulses at 3100\,nm wavelength are presented in Fig.~\ref{fig1}(a). It is interesting to compare our results to earlier ATI experiments. Despite the long laser wavelength and owing to the relatively low intensity, the ponderomotive energy in our experiments amounts to only few eV. This is much less than in previous ATI experiments using mid-IR light, where rare gases or molecules were used as targets, e.g.~\cite{Colosimo2008, Wolter2015, Fuest2019}. The low energy of recolliding electrons counteracts the infamous decay of the recollision probability \cite{Colosimo2008} such that we observe a pronounced ATI plateau. Indeed, the situation in our experiment is rather comparable to the typical case of xenon ionized by 800\,nm light, e.g.~\cite{Paulus2003, Milosevic2006}.

The high-energy cut-off of the ATI plateau at $10\Up$ is used to determine the laser intensity for each measurement. Using these values, we plot the intensity dependent electron yields in Fig.~\ref{fig1}(d). The measured data points agree well with predictions based on the Perelomov-Popov-Terent'ev (PPT) ionization rate \cite{Perelomov1966}. 
The significant flattening of the curve indicates that saturation takes place close to $2\,\mathrm{TW/cm^2}$. In addition, the saturation intensity can be estimated using numerical solutions of the three-dimensional time-dependent Schrödinger equation (3D TDSE). To this end, we evaluate the survival probability at the end of the laser pulse, and find that saturation occurs between $2\,\mathrm{TW/cm^2}$ and $3\,\mathrm{TW/cm^2}$, in good agreement with Fig.~\ref{fig1}(a), see Supplementary Material (SM) \footnote{Supplementary Material is available at \emph{<url>} and includes Refs. \cite{Milosevic2006,Milosevic2010,Gregoire2015,Sansonetti2009,Delone1999,Radzig1985,Bunge1993,
Mittleman1959,Green1969,Green1981,Augst1991,Gibson1994,Walker1998,Kopold2002,Becker2002,
Gazibegovic2004,Hasovic2007,Cerkic2009,Milosevic2014,Fetic2020,vanDruten1995,Milosevic2007,
Roudnev2007,Milosevic2016}}.

CEP-dependent experimental results for ATI of Cs are presented in Figs.~\ref{fig1}(b-e). In order to evaluate the CEP-dependence of the ATI spectra we calculate the asymmetry parameter, $A(E,\phi)=\left[R(E,\phi)-L(E,\phi)\right]/\left[R(E,\phi)+L(E,\phi)\right]$. 
Here, $R(E,\phi)$ [$L(E,\phi)$] are the yields of photoelectrons with energy $E$ detected on the right (left) detector for a laser pulse with CEP $\phi$.
The CEP-dependent asymmetry maps of Figs.~\ref{fig1}(b,c) reveal a clear CEP-dependence of the ATI spectrum, including features which are washed out when the data are averaged over the CEP. At low energies where the ATI spectrum is dominated by direct electron emission, the CEP dependence is rather weak and modulations associated with ATI peaks can be seen.
At higher energies, in the recollision plateau, significantly larger asymmetry values are observed and the regions of positive/negative asymmetry are tilted, i.e.~the phase-of-the phase depends on the electron energy in the characteristic fashion of the ATI plateau \cite{Paulus2003}.

The striking feature of our experimental results is shown in Fig.~\ref{fig1}(b) and highlighted by the black box. At an electron energy of $E\sim 20\,$eV, the CEP-dependent asymmetry trifurcates and exhibits a clear high-order oscillation with $m=3$. Around $E\sim 30\,$eV, the asymmetry returns to the usual periodicity. This range corresponds to $5 \Up \lesssim E \lesssim 7 \Up$. The unusual behavior of the CEP-dependent asymmetry is only observed at the highest intensity studied in our experiments, $I = 4\,\mathrm{TW/cm^2}$, while it is absent at lower intensity values, as shown in Fig.~\ref{fig1}(c). The CEP-dependent asymmetry at $E=6\Up$ [black arrows in Fig.~\ref{fig1}(a)] is quantified in Fig.~\ref{fig1}(e). Clearly, only at the highest intensity value, high-order oscillations are observed. 
We note that the intensity of $I = 4\,\mathrm{TW/cm^2}$ is well beyond the saturation intensity of Cs, see Fig.~\ref{fig1}(d).
The role of saturation is additionally corroborated by unpublished experiments on ATI of alkali atoms using $1800\,$nm few-cycle pulses \cite{Zille_phd}. 

\begin{figure}
\includegraphics[width=0.48 \textwidth]{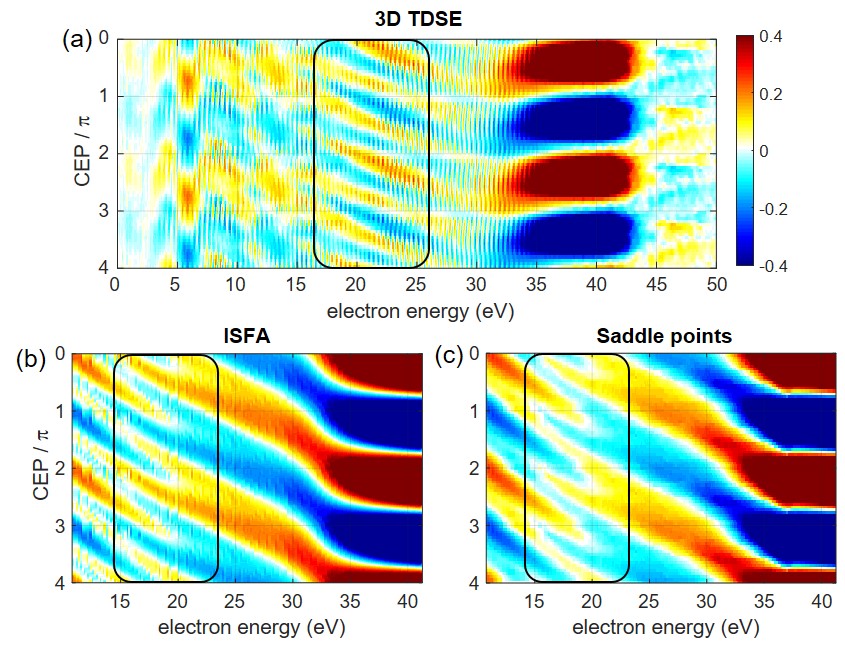}
\caption{Calculated CEP-dependent asymmetry maps for (a) TDSE, (b) ISFA, (c) saddle-point calculations at intensity $I = 4\,\mathrm{TW/cm^2}$, averaged over the intensity distribution of a Gaussian focal volume. While the TDSE results contain contributions from both direct and recollision electrons, the ISFA and saddle-point results are obtained for recolliding electrons only.} \label{fig2}
\end{figure}

In order to interpret our experimental results and find the origin of the asymmetry trifurcation, we employ three different theoretical methods. The most accurate, and also most computationally expensive method is the 3D TDSE, see Ref.~\cite{Fetic2020} and SM for details on the implementation used in the present work. An asymmetry map calculated with the TDSE is displayed in Fig.~\ref{fig2}(a) and agrees very well with the experimental results shown in Fig.~\ref{fig1}(b), despite overestimating the asymmetry amplitude in the cut-off region above 35\,eV. The phase dependence of the asymmetry varies throughout the electron spectrum, not only by phase but also by periodicity. In particular, the high-order asymmetry oscillations around 20-25\,eV can be clearly seen. 

For further analysis, we turn to the ISFA,
\cite{Becker2002,Milosevic2003,Paulus2003,Milosevic2006}, see SM for details.
The ISFA is a quantum-mechanical theory which can be interpreted using the well-known three-step model \cite{Corkum1993}, where electrons first tunnel from the atom around the field maxima, propagate in the continuum
under the sole influence of the laser field, and finally scatter off the parent ion upon recollision. The
electron acquires a phase given by the classical action $S_\mathbf{p}(t,t_0;\phi)$. For a laser pulse with CEP $\phi$, the yield
$y_\mathbf{p}$ of electrons with drift momentum $\mathbf{p}$ can be obtained by integrating over all possible ionization times $t_0$ and
rescattering times $t$,
\begin{equation}
y_\mathbf{p}= p\left|\int dt\int dt_0 V_\mathbf{pk} \left( \frac{2\pi}{i\tau} \right)^{3/2} e^{i S_\mathbf{p}(t,t_0;\phi)}I_{\mathbf{k} b}
\right|^2,\label{eq:sfa}
\end{equation}
where $\tau=t-t_0$ is travel time, $V_\mathbf{pk}$ is the Fourier transform of the rescattering potential $V$ ($\mathbf{k}$ is the
intermediate electron momentum), and $I_{\mathbf{k} b}=\langle\mathbf{k}+\mathbf{A}(t_0)|\mathbf{r}\cdot\mathbf{E}(t_0)|\psi_b\rangle$ is the
ionization dipole matrix element ($\psi_b$ is the bound ground-state wave function).

An asymmetry map calculated by numerically integrating Eq.~(\ref{eq:sfa}) is displayed in Fig.~\ref{fig2}(b). We focus on the region of the rescattering plateau, where we observe a pronounced oscillation with $m=3$ in the energy range up to $\sim 25\,$eV, where it returns to $m=1$, in reasonable agreement with the experimental data. Since direct electrons are neglected in the ISFA results, the low-energy part cannot be directly compared to the experimental results and is omitted.

It is insightful to evaluate the above integral [Eq.~(\ref{eq:sfa})] using the saddle-point approximation. The condition of stationary action yields a discrete number of contributions to the total integral. These contributions are analogous to classical trajectories and referred to as quantum orbits \cite{Paulus2001,Milosevic2006}. Each quantum orbit corresponds to ionization and recollision at different times, $t_0$, $t$, respectively. 
The saddle-point results based on 8 pairs of quantum orbits are displayed in Fig.~\ref{fig2}(c). They match the results of the full numerical integration of Fig.~\ref{fig2}(b) very closely, indicating that all significant contributions to the relevant photoelectron yield are included by considering a limited number of quantum orbits. In the following, we will interpret the CEP-dependent asymmetry and its trifurcation in terms of quantum orbits.

\begin{figure}
\includegraphics[width=0.48 \textwidth]{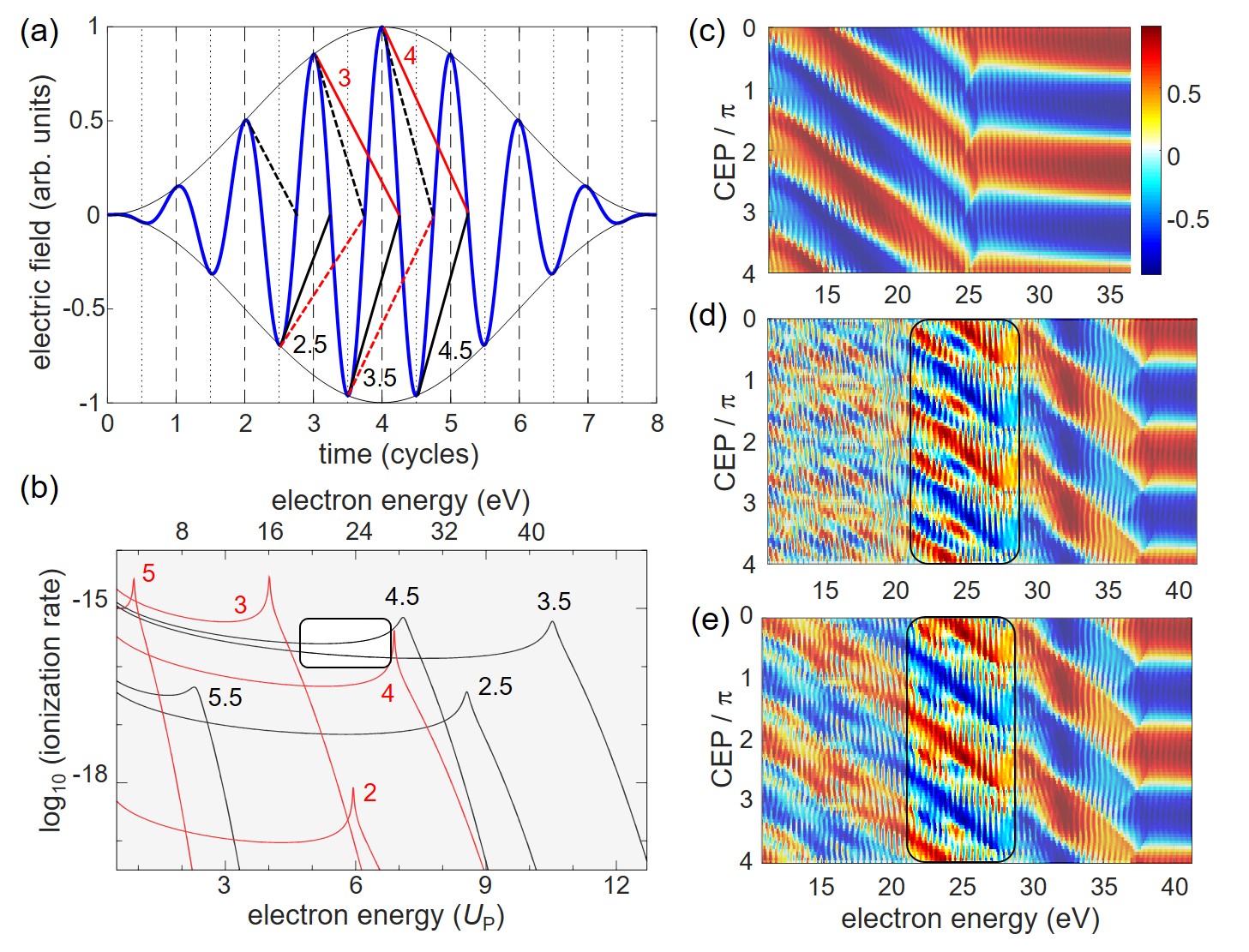}
\caption{Quantum orbit analysis. (a) The laser electric field of a cos-like pulse ($\phi = 0$) is plotted along with the most relevant quantum orbits where the emission time $t_0$ is connected with the recollision time $t$. The solid (dashed) lines correspond to quantum orbits with electron emission in positive (negative) direction. Red integer (black half-integer) values correspond to long (short) orbits that lead to positive drift momenta. (b) For each quantum orbit, the electron yield is plotted against the electron kinetic energy. The sharp spikes at the cut-off energies are an artefact of the saddle-point method. For the intensity of $4\,\mathrm{TW/cm^2}$, the resulting CEP-dependent asymmetry map is plotted for (c) long orbits only, (d) short orbits only, and (e) all orbits. The black boxes mark the trifurcation region.} \label{fig3}
\end{figure}

Using the saddle-point method we obtain quantum orbits that are associated with wavepackets that are created at specific half-cycle maxima of the electric field, at times $t_0$, and recollide at later times $t$. Based on the travel times $\tau = t-t_0$, we distinguish between short orbits with $\tau \approx 0.7 T$ ($T=10.5\,$fs being an optical cycle at $\lambda = 3100\,$nm) that rescatter on the first return, and long orbits with $\tau \approx 1.2 T$ that rescatter on the second return. Orbits with longer travel times are not considered. In Fig.~\ref{fig3}(a), we plot $t_0$, and $t$ for the most important orbits in a cos-like pulse ($\phi = 0$). The time $t_0$ can be the birth time of short and long orbits, which produce electron wave packets with opposite drift momenta. For nomenclature, we concentrate on electron emission with positive drift momentum and use the emission time $t_0$ at $\phi = 0$ as labels for different quantum orbits [see Fig.~\ref{fig3}(a)]. Thus, half-integer values refer to short orbits, and full integers refer to long orbits that produce electrons with positive drift momentum. On the other hand, long orbits starting at half-integer times, and short orbits starting at full integer times result in negative drift momentum. These orbits are not labelled and are indicated by dashed lines in Fig.~\ref{fig3}(a). When the CEP is scanned from $0$ to $2\pi$, the emission and rescattering times shift to earlier times, e.g.~orbit 3.5 starts at $t_0 \approx 3T$ for $\phi = \pi$.

Fig.~\ref{fig3}(b) shows that the different orbits contribute to different parts of the spectrum. For example, the short orbit $3.5$ dominates the high energy part ($E \gtrsim 7\Up$) of the electron spectrum. On the other hand, the long orbit $3$ dominates the low energy part up to approximately $4 \Up$. Beyond $5 \Up$, it quickly drops below the yield from orbits $3.5$ and $4.5$. In the range around $6\Up$, where the trifurcation is observed, the orbits 3.5 and 4.5 yield approximately equal contributions to the photoelectron spectrum. 

Figure \ref{fig3}(c) shows the asymmetry map calculated for the contributions of the long orbits only. It exhibits the usual behavior with $m=1$. This indicates that the long orbits alone are not responsible for the high-order oscillations of the CEP-dependent asymmetry. 

The asymmetry map calculated for the short orbits only is shown in Fig.~\ref{fig3}(d) and exhibits clear high-order oscillations up to approximately $30\,$eV, where it returns to the usual behavior with $m=1$. This transition coincides with the cut-off energy of the orbit $4.5$ ($7 \Up$). The absence of high-order oscillations above $7\Up$ implies that the high-order oscillations around $6\Up$ arise from the interference of the two short quantum orbits $3.5$ and $4.5$. Even faster oscillation are present at energies below $20\,$eV in Fig.~\ref{fig3}(d). However, these cannot be directly observered, since the dominant contribution is given by the long quantum orbit $4$ [see Fig.~\ref{fig3}(b)], which exhibits only the fundamental oscillation, $m=1$.

The full asymmetry map composed of both short and long orbit contributions is displayed in Fig.~\ref{fig3}(e). It can be seen that a pronounced high-order oscillation persists in the range between $20$ and $30\,$eV. At lower energies the behavior of the asymmetry is dictated by the long quantum orbit. One might object that signatures of fast oscillations are still observable at low energies in Fig.~\ref{fig3}(e). However, the same can be said about the experimental data presented in Fig.~\ref{fig1}(b), for example around 15\,eV.
At higher energies ($E>7\Up$), on the other hand, the electron spectrum is dominated by the contributions from a single short quantum orbit 3.5, which also results in the fundamental oscillation period. 

The interference of quantum orbits underlying the observed trifurcation can be interpreted as holography in the time domain. This differs from the position-space holography evoked in previous studies on photoelectron holography, where different trajectories originating in the same laser half-cycle interfere \cite{Huismans2011,Haertelt2016,Walt2017,Porat2018}.
The essential feature of holography is the presence of a signal wave and a reference wave. For a useful reference, one of the interfering orbits in our experiment should exhibit a well-behaved phase evolution. For our case the orbit 3.5 is a good choice for the reference wave; it propagates in the continuum around the center of the pulse, where the intensity variation are small compared to the situation of orbit 4.5, which propagates on the falling edge of the pulse envelope, where intensity variations are significant. For details, see the calculated phase evolution of orbits 3.5 and 4.5 given in the SM.

The question arises why high-order asymmetries have not been reported in previous experiments. Part of the explanation has been implicitly given above: in many experiments, the high-order oscillations are concealed by the dominant contribution of a single quantum orbit. This is the case, in particular, for very short laser pulses, which are otherwise beneficial to observe pronounced CEP effects. In our case, using three-cycle pulses, however, two quantum orbits produce very similar contributions in the energy range between $5\Up$ and $7\Up$, which creates favorable conditions for the high-order oscillations to become observable. In other words, for holographic interferences, the pulse needs to be sufficiently long to allow for one orbit to be considered a reference wave.

The question remains why the effect is only observed at high intensity in our experiments. This can be explained by the effect of ionization depletion during the laser pulse, which suppresses the contribution of later half-cycles with respect to earlier ones. In the highlighted region of Fig.~\ref{fig3}(b), the contribution from orbit 4.5 is stronger than that of orbit 3.5. For a certain degree of depletion the contributions of the two interfering orbits may be equalized, thus maximizing the interference contrast. An additional factor might be that the measurement of high-order oscillations requires both good statistics and excellent CEP stability, which has been challenging to obtain. In the present experiment, these conditions have been met by using a highly CEP-stable high-repetition rate laser \cite{Kurucz2020}.

In conclusion, we have observed high-order CEP effects in ATI of Cs driven by few-cycle MIR laser pulses. Our analysis based on quantum orbit theory shows that the fast oscillations of the CEP-dependent asymmetry can be understood as the interference of two backscattered quantum orbits. At this point it is unclear how this result should be interpreted in terms of the general theory of CEP effects \cite{Roudnev2007}, given the large number of photons ($n>50$) involved. Future experiments could probe nuclear or electronic dynamics by means of time-domain holography of quantum orbits.



\subsection{Acknowledgements}
We thank F. Kohl, A. Rose, T. Weber, F. Ronneberger, and the ELI-ALPS team for technical and logistical support. The authors acknowledge funding by the Deutsche Forschungsgemeinschaft (DFG) in the framework of the Schwerpunktprogramm (SPP) 1840, Quantum Dynamics in Tailored Intense Fields (project 281296000). MK acknowledges funding by the DFG under project no. 437321733. The ELI-ALPS project (GINOP-2.3.6-15-2015-00001) is supported by the European Union and co-financed by the European Regional Development Fund. B.F. and D.B.M. acknowledge support by the Ministry for Education, Science and Youth, Canton Sarajevo, Bosnia and Herzegovina.

\bibliography{trifurcation}


\end{document}